\documentclass[article,12pt]{article}

\usepackage{geometry}
\usepackage{setspace}
\usepackage{graphicx}
\usepackage{amssymb}
\usepackage{amsmath}
\usepackage{mathtools}
\usepackage{amsfonts}
\usepackage{appendix}
\usepackage{color}
\usepackage{multirow,multicol}
\usepackage{booktabs}
\usepackage{pdflscape}

\newcommand{\bbeta}{\boldsymbol{\beta}}

\newcommand{\btheta}{\boldsymbol{\theta}}

\newcommand{\0}{\boldsymbol{0}}

\newcommand{\bC}{\boldsymbol{C}}

\newcommand{\bg}{\boldsymbol{g}}

\newcommand{\bx}{\boldsymbol{x}}
\newcommand{\bX}{\boldsymbol{X}}

\newcommand{\calD}{\mathcal{D}}

\newcommand{\calL}{\mathcal{L}}

\newcommand{\nL}{\textnormal{L}}
\newcommand{\nE}{\textnormal{E}}
\newcommand{\nR}{\textnormal{R}}

\newcommand{\calN}{\mathcal{N}}

\begin{document}

\title{\bf Double shrunken selection operator}

\bigskip

\author{{B. Y\"{u}zba\c{s}{\i}$^1$\footnote{Corresponding author. Email: } and M. Arashi$^2$}
\vspace{.5cm} \\\it$^{1}$Department of Econometrics, Inonu University, Malatya, Turkey
\\\it$^{2}$Department of Statistics, Shahrood University of Technology, Iran
}

\date{}
\maketitle

\begin{quotation}
\noindent {\it Abstract:}
The least absolute shrinkage and selection operator (LASSO) of Tibshirani (1996) is a prominent estimator which selects significant (under some sense) features and kills insignificant ones. Indeed the LASSO shrinks features lager than a noise level to zero. In this paper, we force LASSO to be shrunken more by proposing a Stein-type shrinkage estimator emanating from the LASSO, namely the Stein-type LASSO. The newly proposed estimator proposes good performance in risk sense numerically. Variants of this estimator have smaller relative MSE and prediction error, compared to the LASSO, in the analysis of prostate cancer data set.

\par

\vspace{9pt} \noindent {\it Key words and phrases:} Double shrinking; Linear regression model; LASSO; MSE; Prediction error; Stein-type shrinkage estimator

\par

\vspace{9pt} \noindent {\it AMS Classification:} 62G08, 62J07, 62G20\par

\end{quotation}\par

\section{Introduction}
It is well-known that the least squares estimator (LSE) in the linear regression model, is unbiased with minimum variance. However, dealing with sparse linear models, it is deficient from prediction accuracy and/or interpretation. As a remedy, one may use the least absolute shrinkage and selection operator (LASSO) estimator of Tibshirani (1996). It defines a continuous shrinking operation that can produce coefficients that are exactly ``zero" and is competitive with subset selection and ridge regression retaining good properties of both the estimators. LASSO simultaneously estimates and selects the coefficients of a given linear regression model. Recently, Saleh and Raheem (2015) have proposed an improved LASSO estimation technique based on Stein-rule, where they use uncertain prior information on parameters of interest. See Saleh (2006) for a comprehensive overview on shrinkage estimation with uncertain prior information. Saleh and Raheem (2015) illustrated superiority of a set of LASSO-based shrinkage estimators over the classical LASSO estimator. However, in this paper, we have a different look to improve the LASSO.

In this paper, we present a Steinian LASSO-type estimator by double shrinking the features. Specifically, following James and Stein (1961) and Stein (1981), we propose a set of Stein-type LASSO estimators. We will illustrate how the proposed set of estimators perform well compared to the LASSO.
In all comparisons, we use the $\nL_2$-risk measure of closeness, i.e., for any estimator $\widehat\btheta$ of the vector-parameter $\btheta$, the $\nL_2$-loss function is given by $\calL(\btheta;\widehat\btheta)=\|\widehat\btheta-\btheta\|^2$ and the associated $\nL_2$-risk is evaluated by $\nE\left[\calL(\btheta;\widehat\btheta)\right]$.

In what follows, we propose the set of Stein-type LASSO estimators and evaluate the performance of the proposed estimators, compared to the LASSO, via a Monte Carlo simulation study. We further investigate the superiority of the proposed estimators compared to the LASSO using the prostate cancer data set.

\setcounter{equation}{0}
\section{Linear Model and Estimators}
Consider the linear regression model
\begin{equation}\label{eq21}
Y_i=\beta_0+\beta_1x_{1i}+\ldots+\beta_px_{pi}+\epsilon_i=\beta_0+\bx_i^{\rm \top}\bbeta+\epsilon_i,\quad i=1,\ldots,n,
\end{equation}
where $\epsilon_1,\ldots,\epsilon_n$ are i.i.d. random variables with mean $0$ and variance $\sigma^2$.

Without loss of generality, we will assume that the covariates are centered
to have mean $0$ and take $\widehat\beta_0=n^{-1}\sum_{j=1}^n Y_i=\bar Y$ and replace $Y_i$ in \eqref{eq21} by
$Y_i-\bar Y$ to eliminate $\beta_0$. Then, we also assume $\bar Y=0$ to better concentrate on the estimation of $\bbeta=(\beta_1,\ldots,\beta_p)^{\rm \top}$.

Following Knight and Fu (2000), we consider the bridge estimator of $\bbeta$ by minimizing the penalized least squares criterion
\begin{equation}\label{eq22}
\sum_{i=1}^n\left(Y_i-\bx_i^{\rm \top}\bbeta\right)^2+\lambda_n\sum_{j=1}^p|\beta_j|^\gamma,
\end{equation}
for a given $\lambda_n$ with $\gamma>0$.

In consequent study, we only focus on the special case $\gamma=1$, resulting the LASSO of Tibshirani (1996). We will provide some notes about the use of \eqref{eq22} in conclusions.

\subsection{Stein-type LASSO}
Following Stein (1981), we define the following set of general shrinkage estimators emanating from the LASSO estimator as
\begin{equation}\label{general-estimator}
\widehat\bbeta_n^{\rm S}=\widehat\bbeta_n^{\rm L}+\bg(\widehat\bbeta_n^{\rm L}),
\end{equation}
for some smooth and bounded function $\bg:\mathbb{R}^p\to\mathbb{R}^p$.

Clearly, the shrinkage estimator $\widehat\bbeta_n^{\rm S}$ has smaller $\nL_2$-risk than LASSO, for all $\bg(\cdot)$ satisfying the following inequality
\begin{equation}\label{condition1}
\|\bg(\widehat\bbeta_n^{\rm L})\|^2+2\nabla^{\rm \top}\bg(\widehat\bbeta_n^{\rm L})<0,\quad \textnormal{almost everywhere in}\;\bg.
\end{equation}

Let define $a=(n-p)(p-2)/(n-p+2)$, $\mathcal{W}_n= (\widehat\bbeta_n^{\rm L})^{\rm \top}(\bX^{\rm \top}\bX)\widehat\bbeta_n^{\rm L}/\widehat{\sigma}^2$ and $\widehat{\sigma}^2$ is a consistent estimator of $\sigma^2$ and $\bX=\left( \bx_{1},\ldots,\bx_{n}\right)^{\top}$.  A well-known function which satisfies the condition \eqref{condition1} is $\bg(\widehat\bbeta_n^{\rm L})=-a\mathcal{W}_n^{-1}$, giving rise to the Stein-type estimator, for small enough $a$. However, incorporating such function in \eqref{general-estimator}, gives an estimator with undesirable properties. Apparently as soon as $\mathcal{W}_n<a$, the proposed estimator changes the sign of LASSO. On the other hand, the new estimator does not scale LASSO component-wise. Hence, for $\widehat\bbeta_n^{\rm L}=(\widehat\beta_{1n}^{\rm L},\ldots,\widehat\beta_{pn}^{\rm L})^{\rm \top}$, we define the Stein-type LASSO (SL) estimator with form
\topfigrule

\begin{equation}
\widehat\bbeta_n^{\rm SL}=\left(\left\{1-a\mathcal{W}_n^{-1}\right\}\widehat\beta_{jn}^{\rm L}|j=1,\ldots,p\right)^{\rm \top}.
\end{equation}

Assume $\bC_n=\frac1n\sum_{i=1}^n \bx_i\bx_i^{\rm \top}\to\bC$, $\bC$ is a non-negative definite matrix and \linebreak $\frac1n\max_{1\leq i\leq n}\bx_i^{\rm \top}\bx_i\to0$.
Clearly, if $\lambda_n$ is $\sqrt{n}$-consistent, i.e., $\lambda_n=O(\sqrt{n})$, then from Knight and Fu (2000) we have
$\sqrt{n}(\widehat\bbeta_n^{\rm L}-\bbeta)\overset{\mathcal{D}}{\to}\calN_p(\0,\sigma^2\bC^{-1})$ and the $\nL_2$-risk of SL can be obtained using the Stein's identity (1981).

To avoid negative values, the positive part of SL, namely positive rule Stein-type LASSO (PRSL) will be defined as
\begin{equation}
\widehat\bbeta_n^{\rm PRSL}=\left(\left\{1-a\mathcal{W}_n^{-1}\right\}^{+}\widehat\beta_{jn}^{\rm L}|j=1,\ldots,p\right)^{\rm \top},
\end{equation}
where $b^+=\max(0,b)$.

Then, the $\nL_2$-risk difference is given by
\begin{eqnarray*}
\calD_1&=&\nR(\bbeta;\widehat\bbeta_n^{\rm SL})-\nR(\bbeta;\widehat\bbeta_n^{\rm PRSL})\cr
&=&-\sum_j\nE\left[\left\{1-a\mathcal{W}_n^{-1}\right\}^2I\left(\mathcal{W}_n<a\right)\left(\widehat\beta_{jn}^{\rm L}\right)^2\right]\cr
&&+2\sum_j\nE\left[\left\{1-a\mathcal{W}_n^{-1}\right\}I\left(\mathcal{W}_n<a\right)\left(\widehat\beta_{jn}^{\rm L}(\widehat\beta_{jn}^{\rm L}-\beta_j)\right)\right]\cr
&&<0 .
\end{eqnarray*}
Since for values $\mathcal{W}_n<a$, $1-a\mathcal{W}_n^{-1}<0$ and the expected value of a positive random variable is always positive. Hence the positive part of SL has uniformly smaller $\nL_2$-risk compared to SL.


In forthcoming section, we investigate the performance of the PRSL estimator compared to the LASSO, via a Monte Carlo simulation.

\section{Simulation}
In this section we conduct a Monte Carlo simulation study to evaluate the performance of the PRSL with respect to the LASSO of Tibshirani (1996).

We generate the vector of responses from following model:
\begin{equation}\label{sim.mod}
Y_i=\beta_1x_{1i}+\ldots+\beta_px_{pi}+\epsilon_i,  i=1,\ldots,n,
\end{equation}
where $E(\epsilon_{i}|\bx_i)=0$ and $E(\epsilon_{i}^2)=1$. Furthermore, we generated the predictors $x_{ij}$ and errors $\epsilon_{i}$ from  $\mathcal{N}\left(0,1\right)$. We consider the sample size $n \in \left \{50,100\right \}$ and the number of predictor variables $p  \in \left \{10, 20, 30 \right \}$. We also consider the regression coefficients are set $\beta_j=c\sqrt{2\alpha}j^{-\alpha/2}$ with $\alpha=0.1,0.5,1$ for $j=1,\cdots,p$. The larger values of $\alpha$ indicates that the coefficients $\beta_j$ decline more quickly with $j$. Also, the value of $c$ controls the population ${\rm R}^2=c^2/(1+c^2)$, and is selected on a 20-point grid in $[0,{\rm R}^2]$.

The number of simulations is initially varied. Finally, each realization is repeated 1000 times to obtain stable results. For each realization, we calculated the MSE of suggested estimators. All computations were conducted using the software R. 

The performance of an estimator $\widehat\bbeta_n^{\rm \ast}$ was evaluated by
using MSE criterion, scaled by the MSE of LASSO so that the values of relative MSE (RMSE), is given by
\begin{equation}
\textnormal{RMSE}\left( \widehat\bbeta_n^{\rm \ast}\right) =\frac{\textnormal{MSE}\left( \widehat\bbeta_n^{\rm \ast}\right) }{\textnormal{MSE}\left( \widehat\bbeta_n^{\rm L}\right) }.
\label{rmse}
\end{equation}

If the RMSE is less than one, then it indicates performance superior to the LASSO.

\begin{figure}[!h]
\centering
\includegraphics[width=15cm,height=10cm]{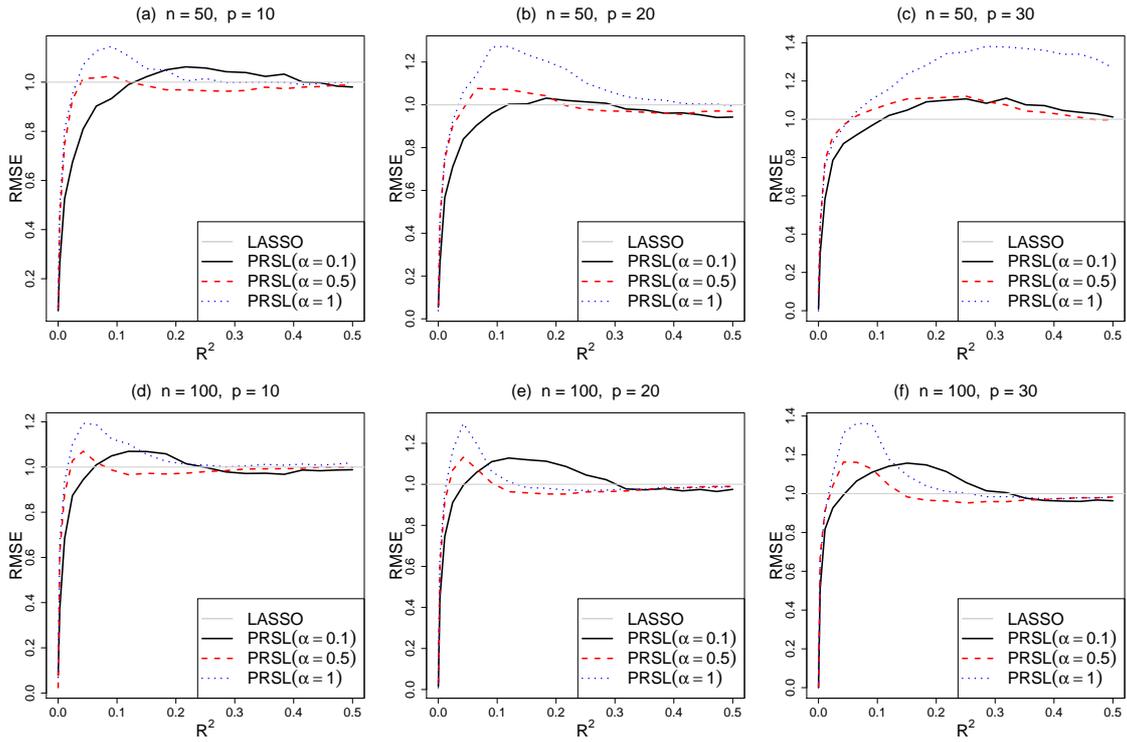}
\caption{ The RMSEs of suggested estimator for different values of $\alpha$ when $\rm R^2\in [0,0.5]$
 \label{Fig:1}}
\end{figure}

\begin{figure}[!h]
\centering
\includegraphics[width=15cm,height=10cm]{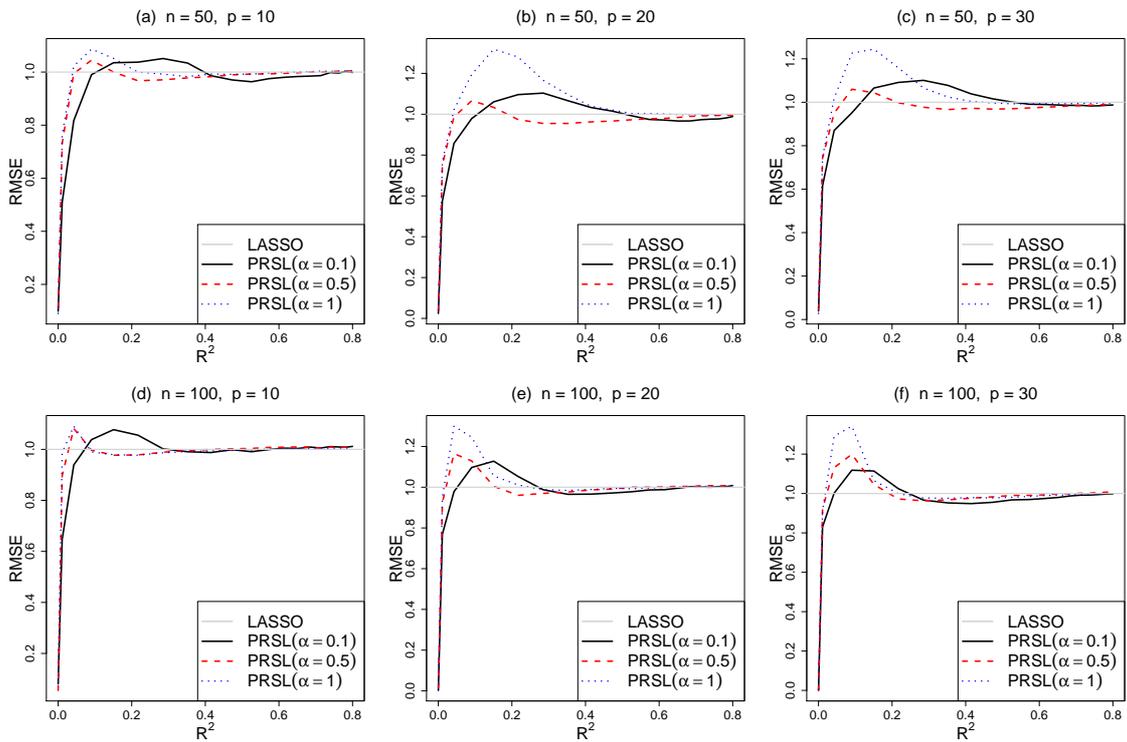}
\caption{ The RMSEs of suggested estimator for different values of $\alpha$ when $\rm R^2\in [0,0.8]$
 \label{Fig:2}}
\end{figure}

The results are reported graphically in Figures~\ref{Fig:1} and ~\ref{Fig:2} for the ease of comparison.
Each figure has six panel plots which correspond to three values of $\alpha$ for $n = 50, 100$ and $p=10,20,30$, and presents the RMSE values of the estimators in Equation \ref{rmse} as a function of the population $\rm R^2$. According to these figures, we can see clear trends. For example, in Figure~\ref{Fig:1}(b), if the $\rm R^2$ varies from 0 to 0.1, then the PRSL has the smallest RMSE when $\alpha=0.1$, which indicates that it performs better than LASSO, followed by the PRSL when $\alpha=0.5$ and $\alpha=1$. On the other hand, for the intermediate values of $\rm R^2$, the performance of PRSL is less efficient than the performance of LASSO. Also, the RMSE of PRSL when $\alpha=0.1,0.5$ is superior to LASSO when $\rm R^2$ is getting increased. If we take a closer look to Figure~\ref{Fig:1}(e), which is the case $(n,p)=(100,20)$, then one can see a similar trend except that the RMSEs of the PRSL outshine the LASSO for each values of $\alpha$ when the population $\rm R^2$ is approaching to $0.5$. In Figure~\ref{Fig:2}, as summary, the performance of PRSL is more efficient than LASSO for the small values of population $\rm R^2$, and it looses its efficiency when we increase in small amounts $\rm R^2$, and finally the relative performance of all estimators become almost similar when $\rm R^2$ is close to 0.8.


\section{Prostate Data}
Prostate data came from the study of Stamey et al. (1989) about correlation between the level of prostate specific antigen (PSA), and a number of clinical measures in men who were about to receive radical prostatectomy. The data consist of 97 measurements on the following variables: log cancer volume (lcavol), log prostate weight (lweight), age (age), log of benign prostatic hyperplasia amount (lbph), log of capsular penetration (lcp), seminal vesicle invasion (svi), Gleason score (gleason), and percent of Gleason scores 4 or 5 (pgg45). The idea is to predict log of PSA (lpsa) from these measured variables.

A descriptions of  the variables in this dataset is given in Table \ref{tab:1}.
			
\begin{table}[!htbp]
	\centering
\caption{Discription of the variables of prostate data}\label{tab:1}{%
\begin{tabular}{|l|l|l|}
\hline
Variables & Description &  Remarks\\
\hline
lpsa & Log of prostate specific antigen (PSA) & Response\\
lcavol & Log cancer volume & \\
lweight & Log prostate weight &\\
age & Age  &  Age in years\\
lbph & Log of benign prostatic hyperplasia amount & \\
svi & Seminal vesicle invasion & \\
lcp & Log of capsular penetration & \\
gleason &  Gleason score& A  numeric vector\\
pgg45 & Percent of Gleason scores 4 or 5& \\
\hline
\end{tabular}}
\end{table}

Playing around with the $\bg$ function in (2.4) may give better candidates compared to LASSO. In this section, we further investigated the performance of the following alternatives
\begin{eqnarray}
\widehat\bbeta_n^{\rm SL2}&=&\left(\left\{1-\frac{a}{\mathcal{W}_n+1}\right\}\widehat\beta_{jn}^{\rm L}|j=1,\ldots,p\right)^{\rm \top}\\
\mbox{or}&&\cr
\widehat\bbeta_n^{\rm SL3}&=&\left(\left\{1-\frac{ar\left(\mathcal{W}_n\right)}{\mathcal{W}_n}\right\}\widehat\beta_{jn}^{\rm L}|j=1,\ldots,p\right)^{\rm \top}
\end{eqnarray}
where $r(x)$ is a concave function w.r.t to $x$, i.e., $r(x)=\sqrt{x}$ or $r(x)=\log |x|$. The latter can be viewed as a Baranchik-type estimator.

\begin{table}[ht]
\centering
\caption{Estimation coeffecients of the variables of prostate data}\label{Tab:2}
\begin{tabular}{rrrrrr}
  \hline
 & LASSO & PRSL & SL2 & SL3($r(x)=\sqrt{x}$) & SL3($r(x)=\log |x|$)\\
  \hline
coef & 2.478 & 2.294 & 2.303 & 0.852 & 1.691 \\
  lcavol & 0.472 & 0.437 & 0.438 & 0.162 & 0.322 \\
  lweight & 0.186 & 0.173 & 0.173 & 0.064 & 0.127 \\
  age & 0.000 & 0.000 & 0.000 & 0.000 & 0.000 \\
  lbph & 0.000 & 0.000 & 0.000 & 0.000 & 0.000 \\
  svi & 0.368 & 0.340 & 0.342 & 0.126 & 0.251 \\
  lcp & 0.000 & 0.000 & 0.000 & 0.000 & 0.000 \\
  gleason & 0.000 & 0.000 & 0.000 & 0.000 & 0.000 \\
  pgg45 & 0.000 & 0.000 & 0.000 & 0.000 & 0.000 \\
   \hline
   RPE & 1.000 &0.764& 0.766& 0.705& 0.335\\
   \hline
\end{tabular}
\end{table}

Our results are based on $1000$ case resampled bootstrap samples. Since there is no noticeable variation for larger number of replications, we did not consider further values. The performance of an estimator is evaluated by its prediction error (PE) via 10-fold cross validation (CV) for each bootstrap replicate. In order to easily compare, we also calculated the relative prediction error (RPE) of an estimator with respect to the prediction error of the LASSO. If the RPE of an estimator is larger than one, then its performance is superior to the LASSO. In Table~\ref{Tab:2}, we report both the estimation coefficient and the APEs of the five methods. According to these results, all suggested estimators outperform the LASSO.

\begin{figure}
\centering
\includegraphics[width=14cm,height=10cm]{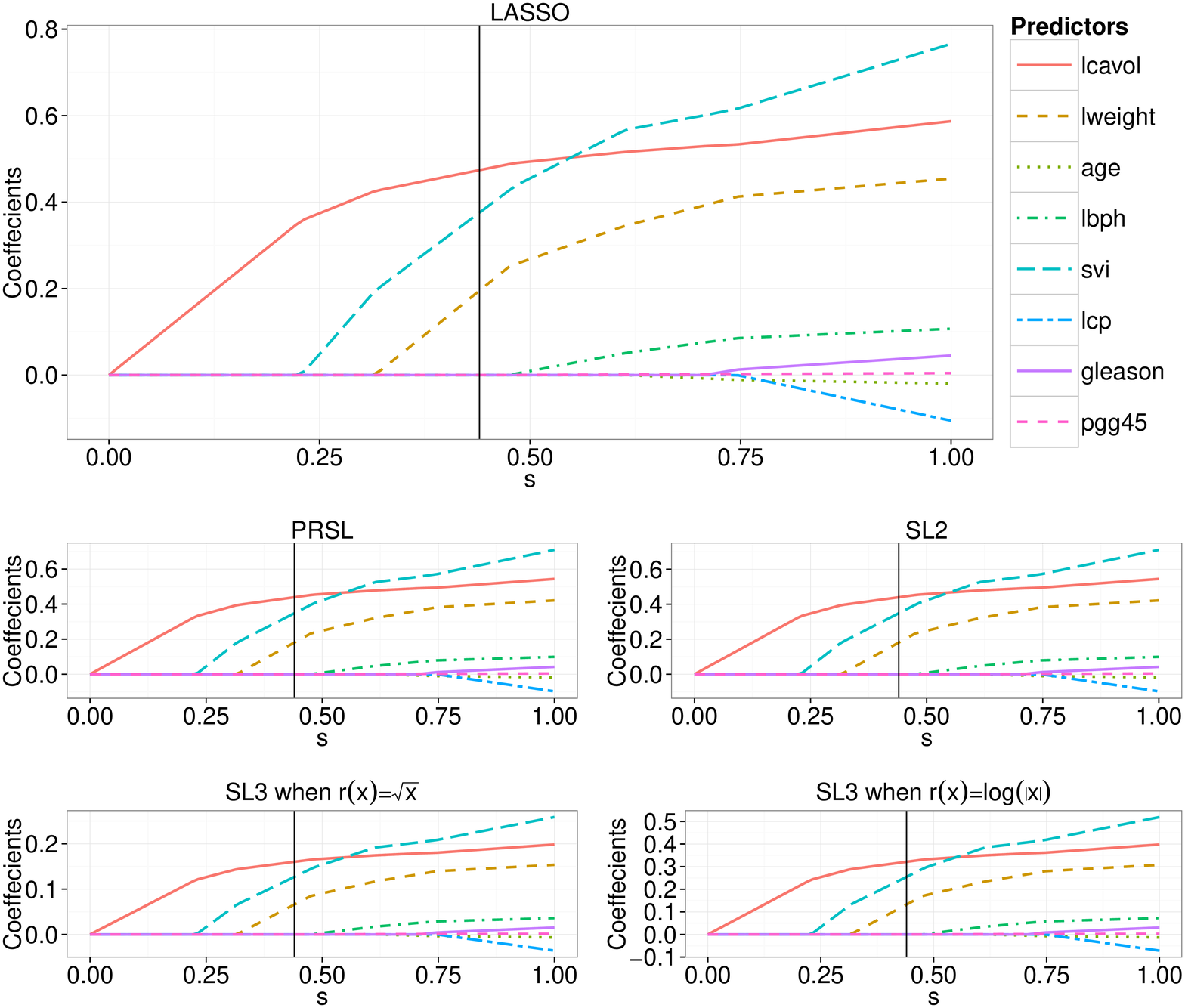}
\caption{ The estimation of coefficients versus $s$ tuning parameter of each methods. Here $s$ is selected via 10-fold CV. The vertical line $\widehat s=0.44$ is selected by ``one standard error" rule.
 \label{Fig:coef:paths}}
\end{figure}

\begin{figure}
\centering
\includegraphics[width=14cm,height=10cm]{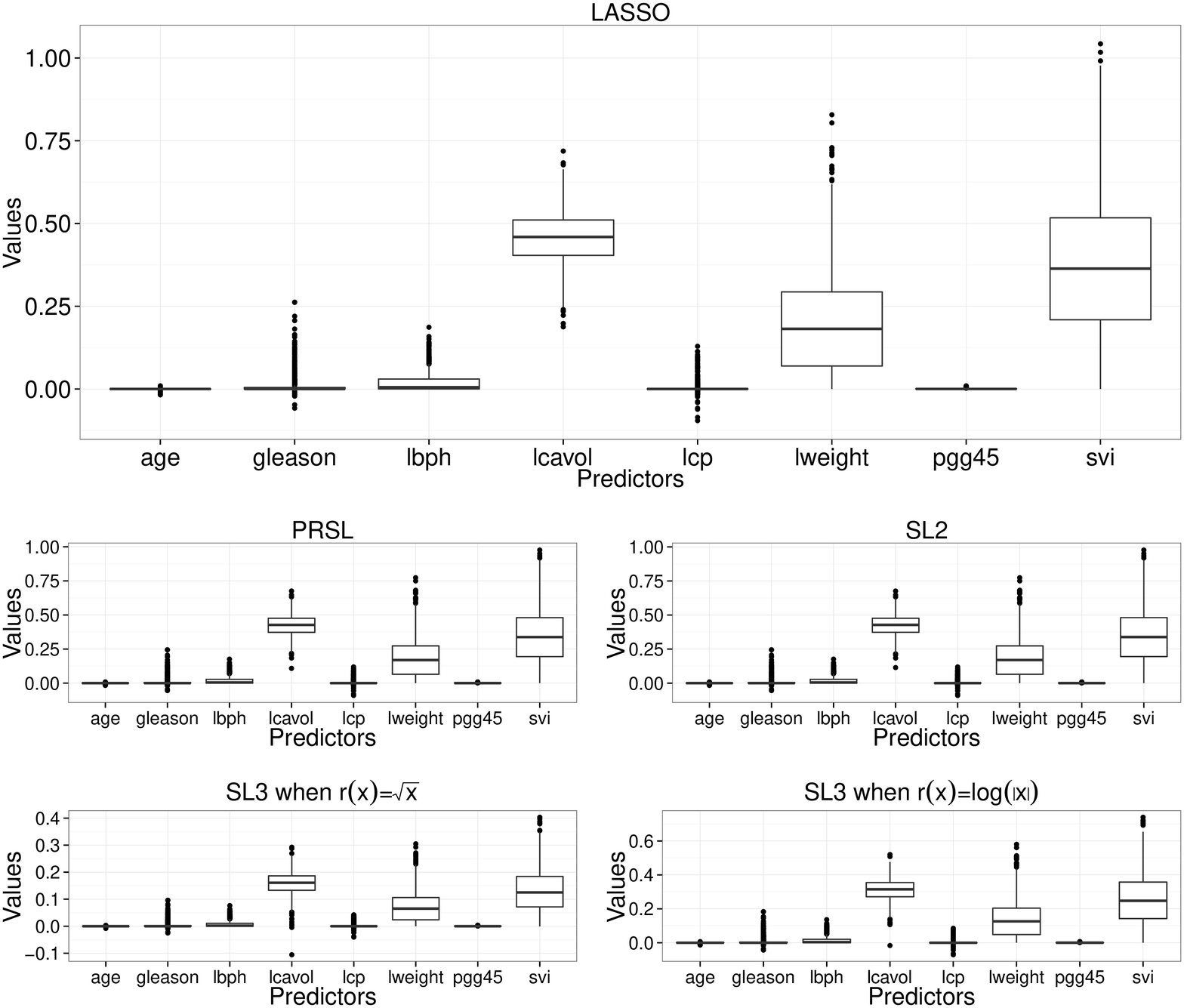}
\caption{ Box plots of 1000 bootstrap values of the listed mothods coefficient estimates for the eight predictors in the prostate cancer example
 \label{Fig:boxplot:prostate}}
\end{figure}

Figure~\ref{Fig:coef:paths} shows each estimates as a function of standardized bound $s=|\bbeta|/max|\bbeta|$. The vertical line represents the model for $\widehat s = 0.44$, the optimal value selected ``one standard error" rule with 10-fold CV, in which we choose the most parsimonious model whose error is no more than one standard error above the error of the best model.  So, all methods gave non-zero coefficients to lcavol, lweight and svi. Also, Figure~\ref{Fig:boxplot:prostate} shows box plots of 1000 bootstrap replications of each methods with $\widehat s=0.44$. And, the results are consistent with Tibshirani (1996).

\section{Conclusions}
In this paper, we employed the shrinkage idea of Stein (1981) to shrink the LASSO of Tibshirani (1996) more. Hence, under the concept of double shrinking, we proposed a double shrinkage estimator namely Stein-type LASSO. Some other similar double shrinkage estimators including the positive part of Stein-type LASSO also proposed as alternative options. Performance analysis of the proposed estimators investigated through a Monte-Carlo simulation as well as a real data analysis. The new set of estimators propose smaller $L_2$-risk compared to the LASSO. Moreover, the prostate cancer data analysis illustrated that the Stein-type LASSO estimators have smaller prediction error compared to the LASSO.

Regarding the function $\bg(\cdot)$ in \eqref{general-estimator}, numerical analysis illustrated that convex and differentiable functions behave superiorly. All our candidates for $\bg(\cdot)$ satisfied the regularity condition \eqref{condition1}. Further, our proposal will also work for the minimizer of \eqref{eq22} for all values $\gamma>0$, including the ridge regression estimator and subset selector. Hence, the proposed methodology can be applied for other estimators. Apart from this, there are many competitors to the LASSO in the context of variable selection, where we only focused on LASSO for the purpose of defining double shrinking idea. For further research, one can use this method to define double shrunken estimator other than the Stein-type LASSO. As such one can define the Stein-type SCAD estimator.

\section*{References}
\baselineskip=12pt
\def\ref{\noindent\hangindent 25pt}



\ref James, W. and Stein, C. (1961). Estimation of quadratic loss, Proc. of the Fourth Berkeley Symp. on Math. Statist. Prob., 1, 361–379.

\ref  Knight K, Fu W. (2000). Asymptotics for LASSO-type estimators. {\em Ann. Statist.}, 10;28(5):1356-1378.

\ref Saleh. A. K. Md. Ehsanes. (2006). Theory of Preliminary Test and Stein-Type Estimation with Applications, Wiley; United Stated of America.

\ref Saleh, A. K. Md. Ehsanes and Raheem, E. (2015). Improved LASSO, arXiv:1503.05160v1, 1-46.

\ref Stamey, T.A., Kabalin, J.N., McNeal, J.E., Johnstone, I.M., Freiha, F., Redwine, E.A. and Yang, N. (1989). Prostate specific antigen in the diagnosis and treatment of adenocarcinoma of the prostate: II. radical prostatectomy treated patients, Journal of Urology 141(5), 1076–1083.

\ref Stein, C. (1981). Estimation of the mean of a multivariate normal distribution, {\em Ann. Statist.} 9, 1135-1151.

\ref Tibshirani, R. (1996). Regression shrinkage and selection via the LASSO, {\em J. Royal. Statist. Soc. B.}, {\bf 58}(1), 267-288.


		

\end{document}